\documentclass{siamltex}
\usepackage{graphicx}
\usepackage{amsmath}
\usepackage[latin1]{inputenc}
\usepackage{t1enc}
\usepackage{amsfonts}
\usepackage{setspace}
\newtheorem{thm}{Theorem}[section]

\newtheorem{lem}[thm]{Lemma}

\newtheorem{rem}[thm]{Remark}

\title{PERIODIC OSCILLATIONS OF BLOOD CELL POPULATIONS IN
CHRONIC MYELOGENOUS LEUKEMIA}
\author{Michael C. Mackey \thanks{Departments of Physiology, Centre for Nonlinear Dynamics, McGill
University, 3655 Drummond, Montr\'eal, Qu\'ebec H3G 1Y6 CANADA, \texttt{ mackey@cnd.mcgill.ca}} \and Chunhua Ou
\thanks{ Laboratory for Industrial and Applied Mathematics Department of Mathematics
and Statistics York University Toronto, Ontario M3J 1P3 CANADA, \texttt{ chqu@mathstat.yorku.ca }}\and
 Laurent Pujo-Menjouet \thanks{
Vanderbilt University, Department of Mathematics, 1326 Stevenson
Center, Nashville, TN 37240-0001, USA,
\texttt{pujo@math.vanderbilt.edu}}
 \and Jianhong Wu, \thanks{
Laboratory for Industrial and Applied Mathematics Department of Mathematics and Statistics York University
Toronto, Ontario M3J 1P3 CANADA, \texttt{ wujh@mathstat.yorku.ca }}}

\begin{document}
\maketitle

\begin{abstract}
We develop some techniques to prove analytically the existence and
stability of long period oscillations of stem cell populations in
the case of periodic chronic myelogenous leukemia. Such a periodic
oscillation $p_\infty $ can be analytically constructed when the
hill coefficient involved in the nonlinear feedback is infinite,
and we show it is possible to  obtain a contractive returning map
(for the semiflow defined by the modeling functional differential
equation) in a closed and convex cone containing $p_\infty $ when
the hill coefficient is large, and the fixed point of such a
contractive map gives the long period oscillation previously
observed both numerically and experimentally.
\end{abstract}

%\doublespace
\baselineskip=14pt
%%% ----------------------------------------------------------------------
%%% ----------------------------------------------------------------------

%-------------------------------------------------------------------------
% ABSTRACT
%-------------------------------------------------------------------------

%-------------------------------------------------------------------------
% FIN DU RESUME

%-------------------------------------------------------------------------
% Keywords
%-------------------------------------------------------------------------


-------------------------------------------------------------------------
% End of keywords


\pagestyle{myheadings} \thispagestyle{plain} \markboth{M. C.
MACKEY, C. OU, L. PUJO-MENJOUET and J. WU}{PERIODIC SOLUTIONS IN
CHRONIC MYELOGENOUS LEUKEMIA}

%**********************************************************
% Introduction
%**********************************************************

\section{Introduction}

\label{sec:introduction} ``How do `short' cell cycles give rise to `long' period oscillations?'' This question
has arisen from the observation of blood cell population oscillations in the case of periodic myelogenous
leukemia (PCML) \cite{fortin}, a blood disease to be discussed in some details below.  Indeed, it has long been
observed in the bone marrow that there is an enormous difference between the relatively short cell cycle
duration which ranges between $1$ to $4$ days \cite{AM2001}, \cite{mackey96} , \cite{MCM2001} and the long
period oscillations in PCML (between $40$ to $ 80$ days) \cite{fortin}. The link between these relatively short
cycle durations and the long periods of peripheral cell oscillations is unclear, to the best of our knowledge,
has neither been biologically explained nor understood. An attempt to answer this question has been made by
Pujo-Menjouet and Mackey in \cite{pujomackey2003} and Pujo-Menjouet et al. in \cite{pujobernardmackey2004},
where they investigated the role of each parameter of the mathematical model involved in the cell cycle and  the
influence of each parameter on the long period and the amplitude of the peripheral cell oscillations. They
showed qualitatively that the cell cycle regulation parameters have major influence on the oscillation amplitude
while the oscillation period is correlated with the cell death and differentiation parameters, and they obtained
these results in the particular case where the hill coefficient involved in the model formulation is infinite.
Our objective here is to prove analytically that the similar conclusions and results remain true in the more
biologically realistic case where the hill coefficient is finite.

More specifically, from the previous studies, it is known that the evolution of the cells in resting phase
involves the Hill function in both the term representing the instantaneous loss of proliferating cells to cell
division and to differentiation and the term representing the delayed production of proliferating stem cells. A
key parameter in the Hill function is the integer $n$ which is usually large, and this Hill function reduces to
the Heaviside step function when $n=\infty$. As will be shown, the underlying system with $n=\infty $ becomes a
piecewise linear scalar delay differential equation that, after non-trivial but straightforward calculations,
has a periodic solution of large periods and amplitudes with very strong stability and attractivity properties.
The main purpose of this paper is to construct a convex closed cone containing the aforementioned periodic
solution (when $n=\infty$) and a contractive returning map defined on this cone such that a fixed point of such
a returning map gives a stable periodic solution of the model equation for the cells in resting phase when $n$
is large. This method was first developed in Walther \cite{walther2001DCDS}, \cite{walther2001AMS} for a scalar
delay differential equation with constant linear instantaneous friction and a negative delayed feedback, and was
later extended to state-dependent delay differential equations \cite{walther2002},\cite {walther2003} and to
delay differential systems \cite{walther2003}, \cite {wu2003}. This method was further enhanced recently in
\cite{OuWU} by incorporating some ideas from classical asymptotic analysis and matching method. Applications of
this method to the model for cells in the resting phase seem to be highly non-trivial since both the
instantaneous loss of proliferating cells and the delayed production of proliferating stem cells involve the
nonlinearity and there is no analytic formula for the periodic solution in the limiting case ($n=\infty$).

We should emphasize that periodic hematological diseases have attracted a significant amount of modelling
attention in various domain such as periodic auto-immune hemolytic anemia \cite {belair1995}, \cite{Mackey79_2}
and cyclical thrombocytopenia \cite {Santillan2000}, \cite{swinburne2000}. It has been observed that the
periodic hematological diseases of this type involve periodicity between two and four times the bone marrow
production delay. This observation has a clear explanation within a modelling context. Some other hematological
diseases such like cyclical neutropenia (\cite{BBM}, \cite{haurie98a}, \cite{HHM98}, \cite{Mackey78},
\cite{Mackey79}, \cite {mackey96}) and chronic myelogenous leukemia \cite{fortin} involve more than one blood
cell type (\textit{i.e.} white cells, red blood cells and platelets). It is believed that the oscillations in
these diseases originate in the pluripotential stem cell compartment and have very long period durations (of
order of weeks to months) in general. In the particular case of the periodic chronic myelogenous leukemia, the
period can range from $40$ to $80$ days and two lines of evidence appear to prove that the oscillations are due
to a destabilization of the stem cell population based in the bone marrow. The first evidence is due to the gene
mutation in the Philadelphia chromosome and responsible for the disease . The mutated cells have been observed
in all the blood cell lineages \cite {BMPLRMW2000}, \cite{EE1997}, \cite{haferlach97}, \cite{jiang97}, \cite
{takahashi98}. The second line of evidence is given in \cite{fortin} where the authors collected clinical data
from the literature, and proved that white blood cells, erythrocytes and platelets oscillate with the same
period.

Periodic chronic myelogenous leukemia (or PCML) takes its name
from the clinical character and the type of leukemia it describes.
Leukemia is a malignant disease characterized by uncontrolled
proliferation of immature and abnormal white blood cells in the
bone marrow, the blood, the spleen and the liver. Its character
can be chronic (in the early stage of myelogenous leukemia) or
acute (in the late stage). The type of cells involved are myeloid,
lymphoid or monocytic depending on the damaged branch of blood
cell production.  The stem cell model by Pujo-Menjouet and Mackey
in \cite{pujomackey2003} and Pujo-Menjouet et al. in \cite
{pujobernardmackey2004} (also called $G_{0}$ model due to the
consideration of the $G_{0}$ resting phase in the cell cycle) is
developed in order to describe the mechanism of the disease under
consideration  \cite{burnstannock}, \cite {mackeydormer82},
\cite{smithmartin}.

The remaining part of this paper is organized as follows. In Section \ref{sec:model} we present the model in
detail. In Section \ref{sec:largen } we recall some previous results from Pujo-Menjouet et al. in
\cite{pujobernardmackey2004} in the case where the Hill coefficient $n$ is infinite. Then, we introduce a more
general result on the perturbed delay equation given in section \ref {sec:perturbed delay equation}, and we
present our main results in section \ref{sec:full asymptotic expansion} including the full asymptotic expansion
for the periodic solutions.

\section{Description of the model}

\label{sec:model}

\begin{figure}[tbp]
\centering
\includegraphics*[width=0.8\linewidth]{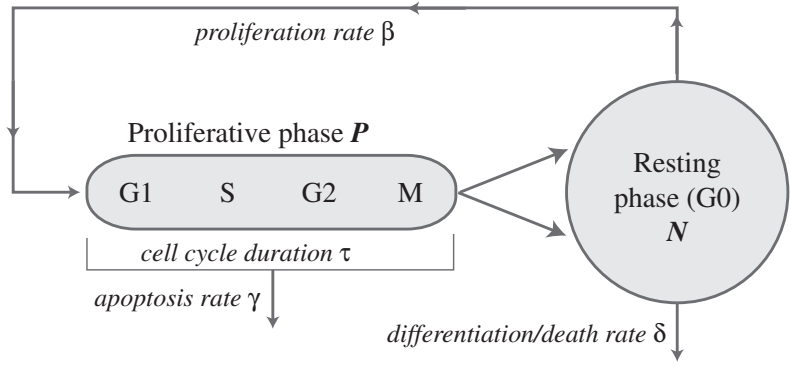}
\caption{A schematic representation of the $G_{0}$ stem cell model.
Proliferating phase cells ($P$) include those cells in $G_{1}$, $S$ (DNA
synthesis), $G_{2}$, and $M$ (mitosis) while the resting phase ($N$) cells
are in the $G_{0}$ phase. $\delta $ is the rate of differentiation into all
the committed stem cell populations, and $\gamma $ represents a loss of
proliferating phase cells due to apoptosis. $\beta $ is the rate of cell
reentry from $G_{0}$ into the proliferative phase, and $\tau $ is the
duration of the proliferative phase. See \protect\cite{Mackey78},
\protect\cite{Mackey79}, \protect\cite{mackey96} for further details.}
\label{cellcycle}
\end{figure}

The $G_{0}$ model, whose early features are due to Lajtha \cite{lajtha1959} and Burns and Tannock
\cite{burnstannock}, is derived from an age structured coupled system of two partial differential equations,
coupled with some boundary and initial conditions \cite{rubinow75}, \cite{Mackey78}, \cite{Mackey79}, and
\cite{MR1}. Using the method of integration along characteristics \cite {webb85} these equations can be
transformed into a pair of non-linear first-order differential delay equations \cite{Mackey78}, \cite{Mackey79},
\cite{mackey96}. The model consists of a proliferating phase where the cell population is denoted by $P(t)$ at
time $t$, and a $G_{0}$ resting phase, with a population of cell $N(t)$. In the proliferating phase, cells are
committed to undergo cell division a constant time $\tau$ after their entry. Note that the choice of $\tau$ as a
constant is to simplify the problem, though some models with a non constant value of $\tau$ exist
\cite{adimypujo2003-2}, \cite{BBM}. The loss rate $\gamma$ in the proliferating phase is due to apoptosis
(programmed cell death). At the point of cytokinesis (cell division), a cell divides into two daughter cells
which enter the resting phase. In this phase, cells can not divide but they have the choice of between three
different fates. They may have one of three possible fates: differentiate at a constant rate $\delta $, reenter
the proliferating phase at a rate $\beta $, or simply remain in $G_0$. Note that the reentering rate $\beta $
will be a nonlinear term in our equation and the focus of our study (see Figure (\ref {cellcycle}) for an
schematic illustration of the cell cycle).

%
%\begin{figure}
%\centering
%\includegraphics* [width=4in,viewport=144 431 432 575]{cellcycle.eps}
%\caption{A schematic representation of the $G_{0}$ stem cell
%model. Proliferating phase cells ($P$) include those cells in
%$G_{1}$, $S$ (DNA synthesis), $G_{2}$, and $M$ (mitosis) while the
%resting phase ($N$) cells are in the $G_{0}$ phase. $\delta $ is
%the rate of  differentiation into all the committed stem cell
%populations, and $\gamma $ represents a loss of proliferating
%phase cells due to apoptosis. $\beta $ is the rate of cell reentry
%from $G_{0}$ into the proliferative phase, and $\tau $ is the
%duration of the proliferative phase. See \cite{Mackey78,
%Mackey79,mackey96} for further details.}\label{cellcycle}
%\end{figure}

The model, described by a coupled non-linear first order delay
equations, takes the following form
\begin{equation}
\frac{dP(t)}{dt}=-\gamma P(t)+\beta (N)N-e^{-\gamma \tau }\beta (N_{\tau
})N_{\tau } ,  \label{w2.1}
\end{equation}
and
\begin{equation}
\frac{dN(t)}{dt}=-[\beta (N)+\delta ]N+2e^{-\gamma \tau }\beta (N_{\tau
})N_{\tau } ,  \label{w2.2}
\end{equation}
where $N_{\tau }=N(t-\tau )$. The resting to proliferative phase
feedback rate $\beta $ is taken to be a Hill function of the form
\[
\beta (N)=\frac{\beta _{0}\theta ^{n}}{\theta ^{n}+N^{n}}.
\]
In equation (\ref{w2.2}), the first term represents the loss of proliferating cells to cell division ($\beta
(N)N$) and to differentiation ($ \delta N$). The second term represents the production of proliferating stem
cells, with the factor $2$ accounting for the amplifying effect of cell division while $e^{-\gamma \tau }$
accounts for the attenuation due to apoptosis. Note that, we only need to understand the dynamics of the $G_{0}$
phase resting population (governed by equation (\ref{w2.2})) since the proliferating phase dynamics (governed by
equation (\ref{w2.1})) are driven by the dynamics of the resting cells.

By introducing the dimensionless variable $x=N/\theta $, we can
rewrite equation (\ref{w2.2}) as
\begin{equation}
\frac{dx}{dt}=-[\beta (x)+\delta ]x+k\beta (x_{\tau} )x_{\tau},  \label{w2.3}
\end{equation}
where
\begin{equation}
\beta (x)=\beta _0\frac 1{1+x^n},  \label{w2.4}
\end{equation}
and $k=2e^{-\gamma \tau }$. The steady state $x_{*}$ of equation (\ref{w2.3} ) are given by the solution of
$dx/dt\equiv 0$. Then we have $x_{*}\equiv 0$ , or
\begin{equation}
x_{*}=\left( \beta _0\frac{k-1}\delta -1\right) ^{1/n}.  \label{w2.5}
\end{equation}
Here we require
\[
\tau <-\frac 1\gamma \ln \dfrac{\delta +\beta _0}{2\beta _0},
\]
so that $\beta _0\dfrac{k-1}\delta >1$ in (\ref{w2.5}).

Note that when $n\rightarrow \infty $, $x_{*}\rightarrow 1$ in
(\ref{w2.5}) and $\beta (x)$ tends to a piecewise constant
function (the Heaviside step function).

A solution of Equation (\ref{w2.3}) is a continuous function $x:[-\tau ,+\infty )\rightarrow \mathbf{R}_{+}$
obeying (\ref{w2.3}) for all $t>0$. The continuous function $\varphi :[-\tau ,0)\rightarrow \mathbf{R}
_{+},\varphi (t)=x(t)$ for all $t\in [-\tau ,0],$ is called the initial condition for $x$. Using the method of
steps, it is easy to prove that for every $\varphi \in C([-\tau ,0]),$ where $C([-\tau ,0])$ is the space of
continuous functions on $[-\tau ,0]$, there is a unique solution of equation (\ref{w2.3}) subject to the initial
condition $\varphi $.

\section{Periodic solutions: limiting nonlinearity}

\label{sec:largen }

In this section we study the dynamics of equation (\ref{w2.3}) when $\beta
(x)$ is reduced to the step function
\[
\beta (x)=\left\{
\begin{array}{c}
0,\;\;x>1, \\
\beta _0,\;x<1.
\end{array}
\right .
\]
As in the paper by Pujo-Menjouet et al. \cite{pujobernardmackey2004}, we
introduce two constants by
\[
\alpha =\beta _0+\delta ,\;\;\Gamma =2\beta _0e^{-\gamma \tau }=k\beta _0.
\]
Inserting the above step function $\beta(x)$ into equation (\ref{w2.3}), we
have \medskip
\begin{equation}  \label{w2.6}
\dfrac{dx}{dt}=\ \left\{
\begin{array}{ll}
-\delta x, & 1\leq x,x_\tau , \\
-\alpha x, & 0\leq x\leq 1\leq x_\tau, \\
-\alpha x+ \Gamma x_\tau, & 0\leq x,x_\tau\leq 1, \\
- \delta x+ \Gamma x_\tau, & 0\leq x_\tau\leq 1\leq x,
\end{array}
\right .
\end{equation}
where $x_\tau=x(t-\tau).$

\medskip
For the equation (\ref{w2.6}), we choose the initial function $ \varphi (t)\geq 1+\eta $ for $t\in [-\tau ,0)$
and $\varphi (0)=1+\eta $ where $\eta $ is a small positive constant chosen later. We should remark  that if we
choose $\varphi (t)\leq 1+\eta $ for $t\in [-\tau ,0)$, the results and the techniques to be obtained and
developed are similar. By the continuity of the solution $x$, we have from equation (\ref{w2.6}) the existence
of a $t_1$ such that $x(t)$ and $ x(t-\tau )$ are greater than $1$ for $t\in [0,t_1)$ and $x(t_1)=1$. The
solution $x(t)$ then satisfies
\begin{equation}
\dfrac{dx}{dt}=-\delta x,\;\;\text{for}\;\;t\in [0,t_1].  \label{w2.7}
\end{equation}
Thus solving the above equation, we can have $x(t)=\varphi
(0)e^{-\delta t}=(1+\eta )e^{-\delta t}$. It follows that
\begin{equation}
t_1=\dfrac{\ln \varphi (0)}\delta =\dfrac{\ln (1+\eta )}\delta .
\label{w2.8}
\end{equation}
In the next interval of time, defined by $[t_1,t_1+\tau ],$ the dynamics are
given by
\begin{equation}
\dfrac{dx}{dt}=-\alpha x.  \label{w2.9}
\end{equation}
The solution is then given by $x(t)=e^{-\alpha (t-t_1)}$ for $t\in
[t_1,t_1+\tau ]$ and satisfies $x(t_1+\tau )=e^{-\alpha \tau }$
which is independent of the initial function $\varphi (t)$. In
other words, the solution destroys all memory of the initial data.

The solution in the next interval will be such that $x,x_\tau <1$.
In order that equation (\ref{w2.6}) has periodic solutions, we
should impose an extra condition on $\Gamma $ and $\alpha $ so
that
\begin{equation}
-\alpha x+\Gamma x_\tau \geq 0.  \label{xx}
\end{equation}
Otherwise, if $-\alpha x+\Gamma x_\tau \leq 0$, then the solution
may tend to zero as $t$ approaches infinity and thus we cannot
expect periodic solution. In particular, if
\[
-\alpha x+\Gamma x_\tau \approx 0,
\]
then the solution may stay below the line of $x=1$ so long that
the analysis becomes very complicated. This is also undesirable
biologically since the period will be extremely long. Note that
for $t\in [t_1+\tau ,t_1+2\tau ]$ , $x(t-\tau )=e^{-\alpha
(t-t_1-\tau )}$. Then, from (\ref{w2.6}), we have
$\dfrac{dx}{dt}=-\alpha x+\Gamma x_\tau =-\alpha x+\Gamma
e^{-\alpha \tau (t-t_1-\tau )}$ which gives
\begin{equation}
x(t)=e^{-\alpha (t-t_1-\tau )}(e^{-\alpha \tau }+\Gamma (t-t_1-\tau )).
\label{w2.10}
\end{equation}
For the sake of simplicity, we impose an extra condition on $ \Gamma $:
\begin{equation}
\Gamma >\max \{\dfrac 1\tau (e^{\alpha \tau }-e^{-\alpha \tau }),\alpha
e^{\alpha \tau }\},  \label{w2.11}
\end{equation}
so that (\ref{xx}) holds when $x\leq 1$ (due to $x(t-\tau )\geq
e^{-\alpha \tau }$), and also $x(t_1+2\tau )=e^{-\alpha \tau
}(e^{-\alpha \tau }+\Gamma \tau )>1$ by (\ref{w2.10}).
\footnote{Note that this condition allows us to get the shortest
period length for the solution. In order to get longer periods, we
should assume other conditions on $\Gamma $ such that $x(t_1+2\tau
)<1$, thus the slope of the increasing part of the solution would
be less steep.}

Since $x(t)$ is increasing in $t\in [t_1+\tau,t_1+2\tau]$, there exists a unique point $t_2\in
(t_1+\tau,t_1+2\tau)$ so that $x(t_2)=1.$ Assume $ t_2=t_1+\tau+u,$ $u\in (0,\tau).$ Then from (\ref{w2.10}) we
have
\begin{equation}
e^{\alpha u}=e^{-\alpha \tau }+\Gamma u.  \label{w2.12}
\end{equation}
The above equation (\ref{w2.12}) is a transcendental equation and cannot be
solved explicitly. But we can expand $e^{\alpha u}$ by Taylor's series, that
is $1+\alpha u+\dfrac{(\alpha u)^2}2$ and solve $u$ by the approximated
equation
\[
1+\alpha u+\dfrac{\alpha ^2}2u^2\approx e^{-\alpha \tau }+\Gamma u.
\]

Next for $t\in [t_2,t_1+2\tau ]$, the dynamics are
\[
\dfrac{dx}{dt}=-\delta x+\Gamma x_\tau =-\delta x+\Gamma e^{-\alpha
(t-t_1-\tau )},
\]
which gives
\begin{equation}
x(t)=e^{-\delta \tau (t-t_2)}\{1-\dfrac \Gamma {\beta _0}e^{\alpha (t_1+\tau
)-\delta t_2}\left( e^{-\beta _0t}-e^{-\beta _0t_2}\right) \}.  \label{w2.14}
\end{equation}
Finally for $t\in [t_1+2\tau ,t_2+\tau ],$
\begin{eqnarray*}
\dfrac{dx}{dt} &=&-\delta x+\Gamma x_\tau , \\
&=&-\delta x+\Gamma e^{-\alpha (t-t_1-2\tau )}(e^{-\alpha \tau }+\Gamma
(t-t_1-2\tau )),
\end{eqnarray*}
that is
\[
x(t)=e^{-\delta (t-t_1-2\tau )}\left[ x(t_1+2\tau )+\Gamma \left(
j(t)-j(t_1+2\tau )\right) \right] ,
\]
where
\[
j(t)=\dfrac 1{(\delta -\alpha )}\left( e^{-\alpha \tau }+\Gamma (t-t_1-2\tau
)-\dfrac \Gamma {\delta -\alpha }\right) e^{(\delta -\alpha )(t-t_1-2\tau
)}.
\]
We now claim that
\begin{equation}
x(t)>1,\,\;t\in (t_2,t_2+\tau ].  \label{w2.15}
\end{equation}
Indeed, at the point $t_2,$ $x(t_2)=1,$ $x(t_2-\tau )\geq e^{-\alpha \tau }.$
By (\ref{w2.11}) we have
\[
x^{\prime }(t_2)>-\delta x+\Gamma x_\tau >0.
\]
Suppose, y way of contradiction, that there exists a point $h\in (t_2,t_2+\tau )$ such that $x(h)=1,$ $x^{\prime
}(h)\leq 0$ and $x(t)>1$ for $t\in (t_2,h).$ Then using equation ( \ref{w2.6}) we have by (\ref{w2.11})
\[
x^{\prime }(h)=-\delta +\Gamma x(h-\tau )\geq -\delta +\Gamma e^{-\alpha
\tau }>0.
\]
This is a contradiction and our claim is true.

After the time $t_2+\tau$, both $x_1$ and $x$ are greater than
$1,$ and the solution satisfies
\begin{equation}
x^{\prime }=-\delta x(t)  \label{w2.16}
\end{equation}
and hence is decreasing. Therefore, there exists a point, say
$t=d$ so that $x(d)=1$. Now we can use (\ref{w2.15}) and
(\ref{w2.16}) to choose a small positive constant $\eta $ such
that at some point $T_x<d$
\begin{equation}
x(T_x)=1+\eta ,\;x(T_x+s)>1+\eta ,\;s\in [-\tau,0).  \label{gh}
\end{equation}
Actually, this $T_x$ is exactly the period of the solution $x(t)$.
Summarizing the above analysis, we have the following result:

\medskip

\begin{thm}
\label{theorem1} Assume that $x$ is the solution of (\ref{w2.6}) subject to the initial condition $\phi \geq
1+\eta $ where $\eta $ is a small positive constant defined in (\ref{gh}). Suppose that $\Gamma $ satisfies (
\ref{w2.11}). Then $x$ is a periodic solution.
\end{thm}

\section{Periodic solutions: general nonlinearity}

\label{sec:perturbed delay equation}

\subsection{Perturbed delay equation}

With the detailed analysis of the $G_0$ model when the Hill
function reduces to the Heaviside step function, we can now
consider the general nonlinearity from the viewpoint of regular
perturbation. More precisely, we consider the perturbed problem
\begin{equation}
\dfrac{dy}{dt}=-[\beta (y)+\delta ]y+k\beta (y_\tau )y_\tau ,  \label{w3.1}
\end{equation}
\textit{i.e.}, we return to the original problem with $\beta
=\beta _0\dfrac 1{1+y^n}$. Denote by $\varepsilon =1/n$, we can
rewrite the Hill function as
\[
\beta (y)=\beta _0\dfrac 1{1+y^{1/\varepsilon }}
\]
As a technical preparation, we now describe some elementary
properties of the above specific Hill function.

\begin{lem}
\label{lemma1} Assume that $\varepsilon $ is sufficiently small. We have

(a) If $y>\left( \dfrac 1\varepsilon \right) ^{\varepsilon /(1-\varepsilon
)},$ then
\[
\beta (y)<\beta _0\varepsilon ,\;\;\;\;y\beta (y)<\beta _0\varepsilon
\]

and if $0<y<\varepsilon ^\varepsilon $, then

\begin{equation}
\beta _0>\beta (y)>\beta _0(1-\varepsilon )\;,\;\;\mathrm{and}\;\;|y\beta
(y)-\beta _0y|<\beta _0\epsilon .  \label{w3.2}
\end{equation}

(b)Moreover,
\[
\left| \dfrac{d(y\beta (y))}{dy}\right| <\beta _0\varepsilon \;,\;\;\mathrm{ for}\;\;y>\left( \dfrac
1\varepsilon \right) ^{2\varepsilon },
\]
and
\[
\left| \dfrac{d(y\beta (y)-\beta _0y)}{dy}\right| <\beta _0\varepsilon
\;,\;\;\mathrm{for}\;\;0<y<(\dfrac{\varepsilon ^2}{1+\varepsilon } )^\varepsilon .
\]
\end{lem}

\textbf{Proof} (a). If $y>\left( \dfrac 1\varepsilon \right) ^{\varepsilon
/(1-\varepsilon )},$ then
\[
\beta (y)=\dfrac{\beta _0}{1+y^{1/\varepsilon }}<\dfrac{\beta _0}{ y^{1/\varepsilon }}<\dfrac{\beta _0}{\left(
\dfrac 1\varepsilon \right) ^{1/(1-\varepsilon )}}<\beta _0\varepsilon,
\]
and
\[
y\beta (y)=\dfrac{\beta _0y}{1+y^{1/\varepsilon }}<\dfrac{\beta _0}{y^{\frac
1\varepsilon -1}}<\beta _0\varepsilon .
\]
If $0<y<\varepsilon ^\varepsilon ,$ then
\[
\beta _0>\beta (y)=\dfrac{\beta _0}{1+y^{1/\varepsilon }}>\beta
_0(1-y^{1/\varepsilon })\geq \beta _0(1-\varepsilon ),
\]
and
\[
|y\beta (y)-\beta _0y|=|\beta _0\dfrac{y^{1/\varepsilon +1}}{ 1+y^{1/\varepsilon }}|< \beta _0 y^{1/\varepsilon
+1}< \beta _0 \varepsilon.
\]

(b) If $y>\left( 1/\varepsilon \right) ^{2\varepsilon },$ then
\[
\left| \left( y\beta (y)\right) ^{\prime }\right| =\beta _0\dfrac{\left|
(\dfrac 1\varepsilon -1)y^{1/\varepsilon }-1\right| }{(1+y^{1/\varepsilon
})^2}\leq \beta _0(\dfrac 1\varepsilon -1)y^{-1/\varepsilon }<\beta
_0\varepsilon .
\]
Since the function
\[
f(x)=\dfrac{(1+\dfrac 1\varepsilon )x+\dfrac 1\varepsilon x^2}{1+x}
\]
is increasing for $x\in (0,\dfrac{\varepsilon ^2}{1+\varepsilon })$ and $f( \dfrac{\varepsilon ^2}{1+\varepsilon
})<\varepsilon ,$ then
\[
\left| \left( y\beta (y)-\beta _0y\right) ^{\prime }\right| =\beta _0\dfrac{ (1+\dfrac 1\varepsilon
)y^{1/\varepsilon }+\dfrac 1\varepsilon y^{2/\varepsilon }}{1+y^{1/\varepsilon }}<\beta _0\varepsilon,
\]
if $0<y<(\dfrac{\varepsilon ^2}{1+\varepsilon })^\varepsilon .$ \endproof

Returning to equation (\ref{w3.1}), we let the initial function $\varphi (t)$
be chosen in the following closed convex sets
\[
A(\eta )=\{\varphi (t)\in C([0,1]):1+\eta \leq \varphi (t)\text{ and } \varphi (0)=1+\eta \},
\]
where $\eta $ is a small positive constant defined in the previous section.
For given $\varphi (t)$ in $A(\eta ),$ we can have a unique solution to
equation (\ref{w3.1}). The relations
\[
F_\beta (t,\varphi )=y_t,\;y_t=y(t+s),\;-\tau \leq s\leq 0,\;t\geq 0
\]
define a continuous semiflow $F=F_\beta $ on $C([-\tau ,0]).$

We find that for the simpler equation (\ref{w2.6}), if $\varphi (t)\in
A(\eta ),$ then the solution will return to $A(\eta )$ after finite time. We
like to know whether or not this situation still happens for equation (\ref
{w3.1}). The study of this point becomes necessary and also important in
order to build a returning map. Fortunately, we have

\medskip

\begin{lem}
Let $y(t)$ be the solution of equation (\ref{w3.1}) with any initial
function $\varphi \in A(\eta ).$ Then
\[
y(t)=x(t)+O(\varepsilon \log \varepsilon ),
\]
for $t\in [0,T_x]$ where $T_x$ is the period of periodic solution $x(t),$
defined in Theorem \ref{theorem1}, to equation (\ref{w2.6}). \label{lemma2}
\end{lem}

\medskip

\textbf{Proof} From (\ref{w3.1}) we know that the solution $y(t)$ is decreasing in $t$ in the right neighborhood
of the starting point $t=0.$ We can further claim that there exist three points $\eta _1,t_1^y,\eta _2,$ $ \eta
_1<t_1^y<\eta _2,$ so that
\begin{equation}
y(\eta _1)=\left( \dfrac 1\varepsilon \right) ^{2\varepsilon },\;\;y(t_1^y)=1,\;y(\eta _2)=\left(
\dfrac{\varepsilon ^2}{1+\varepsilon } \right) ^\varepsilon ,  \label{r1}
\end{equation}
and in the interval $(0,\eta _2),$ the solution $y(t)$ is decreasing.
Indeed, if $y(t)>\left( \dfrac 1\varepsilon \right) ^{2\varepsilon }>\left(
\frac 1\varepsilon \right) ^{\varepsilon /(1-\varepsilon )}$ , and $y(t-\tau
)>\left( \dfrac 1\varepsilon \right) ^{2\varepsilon }>\left( \frac
1\varepsilon \right) ^{\varepsilon /(1-\varepsilon )},$ then by Lemma \ref
{lemma1} we have
\[
\beta (y)y<\beta _0\varepsilon ,\;\;\beta (y_\tau )y_\tau <\beta
_0\varepsilon
\]
and it follows from equation (\ref{w3.1}) that
\begin{eqnarray}
\dfrac{dy}{dt} &=&-(\delta +\beta (y))y+k\beta (y_\tau )y_\tau ,  \nonumber
\\
&<&-\dfrac \delta 2\;\;\text{as }\varepsilon \rightarrow 0,  \label{r3}
\end{eqnarray}
which means that $y(t)$ is decreasing and there exists a point $\eta _1$ such that $y(\eta _1)=\left(
1/\varepsilon \right) ^{2\varepsilon }$ and $ 1+\eta >y(t)>\left( 1/\varepsilon \right) ^{2\varepsilon }$, for
$t\in (0,\eta _1).$ Similarly at the right neighborhood of $\eta _1,$ say $(\eta _1,\eta _1+\tau /2),$we have
$\beta (y_\tau )y_\tau =O(\varepsilon )$ and ( \ref{r3}) still holds. This means the solution is still
decreasing in $t$ and there exist two points $t_1^y,\eta _2,\,\eta _1<t_1^y<\eta _2$ so that
\[
y(t_1^y)=1,\;\;y(\eta _2)=\left( \dfrac{\varepsilon ^2}{1+\varepsilon } \right) ^\varepsilon .
\]
By the Mean--Value Theorem, it is easy to know that
\[
\left| y(\eta _1)-y(\eta _2)\right| \geq \dfrac \delta 2\left| \eta _1-\eta
_2\right|
\]
or equivalently
\[
\eta _2-\eta _1\leq \dfrac 2\delta (y(\eta _1)-y(\eta _2))=O(-\varepsilon
\log \varepsilon ).
\]
Therefore,
\begin{equation}
t_1^y-\eta _1<\eta _2-\eta _1=O(-\varepsilon \log \varepsilon ).  \label{x1}
\end{equation}
Now using again the equations (\ref{w3.1}) and (\ref{w2.6}) for $t\in
[0,\eta _1],$ we have from Lemma \ref{lemma1}
\[
(x-y)^{\prime }=-\delta (x-y)+O(\varepsilon ),
\]
which implies that
\[
|x(t)-y(t)|=O(\varepsilon )
\]
for $t\in [0,\eta _1].$ In particular at the point $t=\eta _1,$
\[
x(\eta _1)=y(\eta _1)+O(\varepsilon )=\left( \dfrac 1\varepsilon \right)
^{\varepsilon /(1-\varepsilon )}+O(\varepsilon )=1+O(-\varepsilon \log
\varepsilon ).
\]
It follows from (\ref{w2.7}) and (\ref{w2.9}) that $t_1,$ defined in (\ref
{w2.8}) satisfies
\[
t_1=\eta _1+O(-\varepsilon \log \varepsilon ),\;
\]
and
\[
t_1=t_1^y+O(-\varepsilon \log \varepsilon ).
\]

For $t\in [\eta _1,\eta _2],$ since both the derivative of $x(t)$ and $y(t)$
are of the order of $O(1)$ and the length of the interval $[\eta _1,\eta _2]$
is of the order of $O(-\varepsilon \log \varepsilon )$, we can conclude that
\begin{equation}
y(t)=x(t)+O(-\varepsilon \log \varepsilon ).  \label{w3.7}
\end{equation}
For $t\in [\eta _2,\tau +\eta _1],$ $y(t-\tau )>\left( 1/\varepsilon \right)
^{2\varepsilon }.$ By Lemma \ref{lemma1}, we still have
\[
y(t-\tau )\beta (y(t-\tau ))=O(\varepsilon ).
\]
By equation (\ref{w3.1}) we know that the solution $y(t)$ will still be
decreasing for $t\in [\eta _2,\tau +\eta _1].$ Note that $0<y(t)<y(\eta
_2)=\varepsilon ^\varepsilon $, so that (\ref{w3.2}) in Lemma \ref{lemma1}
holds. Thus we can derive from (\ref{w3.1}) that
\begin{equation}
y^{\prime }(t)=-\alpha y(t)+O(\varepsilon ),  \label{w3.8}
\end{equation}
for $t\in [\eta _2,\tau +\eta _1].$ Coupling this equation with (\ref{w2.9})
and using (\ref{w3.7}) at the point $t=\eta _2$ gives
\[
y(t)=x(t)+O(\varepsilon \log \varepsilon )
\]
for $t\in [\eta _2,\tau +\eta _1].$

For $t\in [\tau +\eta _1,\tau +\eta _2],$ using again the fact that both the
derivatives of $x(t)$ and $y(t)$ are bounded by $O(1)$ and the length of
this interval is of order $O(-\varepsilon \log \varepsilon ),$ we have
\[
y(t)=x(t)+O(\varepsilon \log \varepsilon ).
\]

For $t\geq 1+\eta _2,$ the solution $y(t)$ begins to increase since $\Gamma $
satisfies (\ref{w2.11}). By the similar argument used above, it follows that
there exist three point $\eta _3,t_2^y,$ $\eta _4,\eta _3<t_2^y<$ $\eta _4$
such that
\[
y(\eta _3)=\left( \dfrac{\varepsilon ^2}{1+\varepsilon }\right) ^\varepsilon
, \; \; \; y(t_2^y)=1, \; \; \; y(\eta _4)=\left( \dfrac 1\varepsilon
\right) ^{2\varepsilon }\;,
\]
and
\begin{equation}
\eta _3=t_2^y+O(\varepsilon \log \varepsilon ),\; \; \; \eta
_4=t_2^y+O(\varepsilon \log \varepsilon ) ,  \label{x3}
\end{equation}
and
\begin{equation}
t_2^y=t_2+O(\varepsilon \log \varepsilon ).  \label{x3a}
\end{equation}
We can continue the process above and it is not difficult to find that $y(t)$
will satisfy
\[
y(t)=x(t)+O(\varepsilon \log \varepsilon ),
\]
for $t\in [0,\eta _4].$

By (\ref{w2.11}) and (\ref{w3.1}) we find that the solution $y(t)$ is
increasing at the point $t=\eta _4$ and in the interval $[\eta _4,\tau +\eta
_3],$ $y(t)$ and $x(t)$ satisfy
\[
(x-y)^{\prime }=-\delta (x-y)+O(\varepsilon ),
\]
and
\[
(x-y)|_{\eta _4}=O(\varepsilon \log \varepsilon ).
\]
So we have
\[
x-y=O(\varepsilon \log \varepsilon ),\text{ for }t\in [\eta _4,\tau +\eta
_3].
\]
For $t\in [\tau +\eta _3,\tau +\eta _4],$ using the same argument as in the
interval $[\tau +\eta _1,\tau +\eta _2],$ we have
\begin{equation}
y(t)=x(t)+O(\varepsilon \log \varepsilon ).  \label{t1}
\end{equation}
Finally for $t\geq \tau +\eta _4,$ the solution is decreasing and attains
the value $1+\eta $ at some point $T_y.$ To be more specific, we have
\begin{equation}
y(t)^{\prime }=-\delta y(t)+O(\varepsilon ),  \label{t2}
\end{equation}
and
\begin{equation}
x(t)=-\delta x(t).  \label{t3}
\end{equation}
Using (\ref{t1}), (\ref{t2}) and (\ref{t3}) we can derive that
\begin{equation}
y(t)=x(t)+O(\varepsilon \log \varepsilon ),\;  \label{t4}
\end{equation}
and
\begin{equation}
T_y=T_x+O(\varepsilon \log \varepsilon ).  \label{t5}
\end{equation}
Furthermore, we also have $y(T_y)=1+\eta $ and
\begin{equation}
y(t)\geq 1+\eta \text{ for }[T_y-\tau ,T_y]  \label{t6}
\end{equation}
provided that $\varepsilon $ is sufficiently small.\endproof

\begin{rem}
By Lemma \ref{lemma2} and equation (\ref{w3.1}) we can have two positive
constants $M_1$ and $M_2$ which are independent of $\varepsilon ,$ so that
\begin{equation}
|y(t)|\leq M_1,  \label{wm1}
\end{equation}
and
\begin{equation}
\left| \dfrac{dy}{dt}\right| \leq M_2.  \label{wm2}
\end{equation}
\end{rem}

Now we are ready to define a continuous returning map
\[
R:A(\eta )\ni \varphi \rightarrow y_{q(\varphi )}=F_\beta (q(\varphi
),\varphi )\in A(\eta ),
\]
where $q(\varphi )=T_y.$ In order to verify that there exists a unique fixed
point in $A(\eta )$ for this map $R$, we need to derive its Lipschitz
constant estimation and show this map $R$ is contractive, i.e., its
Lipschitz constant is less than 1.

\subsection{ Lipschitz constant for the map R}

Lipschitz constants of maps $T:D_T\rightarrow Y,$ $D_T\subset X,$ $X$ and $Y$
normed linear space, are given by
\[
L(T)=\sup_{x\in D_T,y\in D_T,x\neq y}\dfrac{||T(x)-T(y)||}{||x-y||}.
\]
In the case when $D_T=X=\mathbf{R}$ and $\sigma =[x_1,x_2]\in \mathbf{R},$
and $f=T$ we set
\[
L_{[x_1,x_2]}(f)=L(f|[x_1,x_2]).
\]

In the case when $f=y\beta (y),$ we define the following four Lipschitz
constants

\[
\begin{array}{ll}
L_1= & L_{[1+\eta ,+\infty )}(y\beta (y)), \\
&  \\
L_2= & L_{[(\dfrac 1\varepsilon )^{2\varepsilon },+\infty )}(y\beta (y)), \\
&  \\
L_3= & L_{(0,+\infty )}(y\beta (y)) , \\
&  \\
L_4= & L_{\left( 0,\left( \dfrac{\varepsilon ^2}{1+\varepsilon }\right)
^\varepsilon \right) }(y\beta (y)).
\end{array}
\]

Similarly for the function $f=y\beta (y)-\beta _0y,$ we also define the
following Lipschitz constant for later use,
\[
L_5=L_{\left( 0,\left( \dfrac{\varepsilon ^2}{1+\varepsilon }\right)
^\varepsilon \right) }(y\beta (y)-\beta _0y).
\]

\medskip

\begin{thm}
\label{theorem2} When $\varepsilon $ is small, the Lipschitz constant for
the map $R$ satisfies
\[
\lim_{\varepsilon \rightarrow 0}L_R=0<1
\]
\end{thm}

\textbf{Proof} Step 1. For $\phi ,\bar{\phi}$ in $A(\eta ).$ By a similar manner as in the proof of Lemma
\ref{lemma2}, we conclude that there exist $ \eta _1,\eta _2$ and $\bar{\eta}_1,\bar{\eta}_2$ such that,
respectively,
\[
y^\phi (\eta _1)=\left( \dfrac 1\varepsilon \right) ^{2\varepsilon
},\;y^\phi (\eta _2)=(\dfrac{\varepsilon ^2}{1+\varepsilon })^\varepsilon
,\;\eta _1-\eta _2=O(-\varepsilon \log \varepsilon, )
\]
and
\[
y^{\bar{\phi}}(\bar{\eta}_1)=\left( \dfrac 1\varepsilon \right) ^{2\varepsilon
},\;y^{\bar{\phi}}(\bar{\eta}_2)=(\dfrac{\varepsilon ^2}{ 1+\varepsilon })^\varepsilon
,\;\bar{\eta}_1-\bar{\eta}_2=O(-\varepsilon \log \varepsilon ).
\]
Let
\[
\eta _{\min }=\min \{\eta _1,\bar{\eta}_1\},
\]
and
\[
\eta _{\max }=\max \{\eta _2,\bar{\eta}_2\}.
\]
Then by (\ref{x1}) we have
\begin{equation}
\eta _{\max }-\eta _{\min }=O(\varepsilon \log \varepsilon ).  \label{w3.9}
\end{equation}

For $t\in [0,\eta _{\min }],$ using the equation (\ref{w3.1}) for $y^\phi
(t) $ and $y^{\bar{\phi}}(t)$ , respectively, gives
\begin{equation}
\dfrac{dy^\phi (t)}{dt}=-[\delta +\beta (y^\phi (t))]y^\phi (t)+k\beta
(y^\phi (t-\tau ))y^\phi (t-\tau ),  \label{w3.10}
\end{equation}
and
\begin{equation}
\dfrac{dy^{\bar{\phi}}(t)}{dt}=-[\delta +\beta (y^{\bar{\phi}}(t))]y^{\bar{ \phi}}(t)+k\beta
(y^{\bar{\phi}}(t-\tau ))y^{\bar{\phi}}(t-\tau ). \label{w3.11}
\end{equation}
Now we begin to estimate the difference between $y^\phi (t)$ and $y^{\bar{ \phi}}(t).$ Coupling with
(\ref{w3.10}) and (\ref{w3.11}) yields
\begin{eqnarray}
(y^\phi -y^{\bar{\phi}})^{\prime } &=&-\delta (y^\phi -y^{\bar{\phi}})
\label{w3.12} \\
&&-[\beta (y^\phi )y^\phi -\beta (y^{\bar{\phi}})y^{\bar{\phi}}]  \nonumber
\\
&&+k[\beta (y_\tau ^\phi )y_\tau ^\phi -\beta (y_\tau ^{\bar{\phi}})y_\tau ^{ \bar{\phi}}].  \nonumber
\end{eqnarray}
Substituting the following inequalities
\[
|\beta (y^\phi )y^\phi -\beta (y^{\bar{\phi}})y^{\bar{\phi}}|\leq L_2|y^\phi
-y^{\bar{\phi}}|,
\]
and
\[
|\beta (y_\tau ^\phi )y_\tau ^\phi -\beta (y_\tau ^{\bar{\phi}})y_\tau ^{ \bar{\phi}}|\leq L_1||\phi
-\bar{\phi}||
\]
into (\ref{w3.12}), we have
\begin{equation}
(y^\phi -y^{\bar{\phi}})^{\prime }\leq \left( \delta +L_2\right) |y^\phi -y^{ \bar{\phi}}|+kL_1||\phi
-\bar{\phi}||.  \label{w3.14}
\end{equation}
Integrating (\ref{w3.14}) from $0$ to $t,$ gives
\[
(y^\phi -y^{\bar{\phi}})\leq \int_0^t\left( \left( \delta +L_2\right)
|y^\phi -y^{\bar{\phi}}|+kL_1||\phi -\bar{\phi}||\right) ds.
\]
Similarly we have
\[
-(y^\phi -y^{\bar{\phi}})\leq \int_0^t\left( \left( \delta +L_2\right)
|y^\phi -y^{\bar{\phi}}|+kL_1||\phi -\bar{\phi}||\right) ds.
\]
Thus,
\begin{equation}
|y^\phi -y^{\bar{\phi}}|\leq \int_0^t\left( \left( \delta +L_2\right)
|y^\phi -y^{\bar{\phi}}|+kL_1||\phi -\bar{\phi}||\right) ds.  \label{gr}
\end{equation}
Solving (\ref{gr}) ( or by Gronwall inequality), we obtain
\begin{equation}
|y^\phi -y^{\bar{\phi}}|\leq C_1||\phi -\bar{\phi}||,  \label{w3.15}
\end{equation}
where
\begin{equation}
C_1=\dfrac{e^{(\delta +L_2)\eta _{\min }}-1}{\delta +L_2}kL_1.  \label{w3.16}
\end{equation}

Step 2. Next for $t\in [\eta _{\min },\eta _{\max }],$ we have
\[
|\beta (y^\phi )y^\phi -\beta (y^{\bar{\phi}})y^{\bar{\phi}}|\leq L_3|y^\phi
-y^{\bar{\phi}}|,
\]
and
\[
|\beta (y_\tau ^\phi )y_\tau ^\phi -\beta (y_\tau ^{\bar{\phi}})y_\tau ^{ \bar{\phi}}|\leq L_1||\phi
-\bar{\phi}||.
\]
Thus from (\ref{w3.10}) and (\ref{w3.11}) we can obtain as before
\[
|y^\phi -y^{\bar{\phi}}|\leq \int_{\eta _{\min }}^t\left( \left( \delta
+L_3\right) |y^\phi -y^{\bar{\phi}}|+kL_1||\phi -\bar{\phi}||\right)
ds+C_1||\phi -\bar{\phi}||.
\]
Then by Gronwall inequality we have
\begin{equation}
|y^\phi -y^{\bar{\phi}}|\leq C_2||\phi -\bar{\phi}||  \label{w3.17}
\end{equation}
where
\begin{equation}
C_2=C_1e^{(\delta +L_3)(\eta _{\max }-\eta _{\min })}+\dfrac{e^{\delta
+L_3(\eta _{\max }-\eta _{\min })}-1}{\delta +L_3}kL_1>C_1.  \label{w3.18}
\end{equation}

Step 3. For $t\in [\eta _{\max },\tau +\eta _{\min }],$
\[
|\beta (y^\phi )y^\phi -\beta _0y^\phi -(\beta (y^{\bar{\phi}})y^{\bar{\phi} }-\beta _0y^{\bar{\phi}})|\leq
L_5|y^\phi -y^{\bar{\phi}}|
\]
and
\[
|\beta (y_\tau ^\phi )y_\tau ^\phi -\beta (y_\tau ^{\bar{\phi}})y_\tau ^{ \bar{\phi}}|\leq L_2||\phi
-\bar{\phi}||.
\]
It is thus easy to have
\[
|y^\phi -y^{\bar{\phi}}|\leq \int_{\eta _{\max }}^t\left( (\alpha
+L_5)|y^\phi -y^{\bar{\phi}}|+kL_2C_1||\phi -\bar{\phi}||\right)
ds+C_2||\phi -\bar{\phi}||,
\]
and to conclude that (due to $1+\eta _{\min }-\eta _{\max }<\tau $)
\begin{equation}
|y^\phi -y^{\bar{\phi}}|\leq C_3||\phi -\bar{\phi}||,  \label{w3.19}
\end{equation}
where
\begin{equation}
C_3=C_2e^{\alpha \tau +\tau L_5}+\dfrac{e^{\alpha \tau +\tau L_5}-1}{\alpha
+L_5}kL_2C_1>C_2.  \label{w3.20}
\end{equation}

Step 4. When $t\geq 1+\eta _{\min },$ We note that from the proof of Lemma
\ref{lemma2}, there exist $\eta _3<\eta _4$, and $\bar{\eta}_3<\bar{\eta}_4$
so that
\[
y^\phi (\eta _3)=(\dfrac{\varepsilon ^2}{1+\varepsilon })^\varepsilon
,\;y^\phi (\eta _4)=\left( \dfrac 1\varepsilon \right) ^{2\varepsilon
}\;,\;\eta _4-\eta _3=O(-\varepsilon \log \varepsilon )
\]
and
\[
y^{\bar{\phi}}(\bar{\eta}_3)=(\dfrac{\varepsilon ^2}{1+\varepsilon } )^\varepsilon
,\;y^{\bar{\phi}}(\bar{\eta}_4)=\left( \dfrac 1\varepsilon \right) ^{2\varepsilon
},\;\;\bar{\eta}_4-\bar{\eta}_3=O(-\varepsilon \log \varepsilon ).
\]
Let
\[
\eta _{\min }^3=\min \{\eta _3,\bar{\eta}_3\},\;\eta _{\max }^4=\max \{\eta
_4,\bar{\eta}_4\}.
\]
Then by (\ref{x3}) we have
\begin{equation}
\eta _{\max }^4-\eta _{\min }^3=O(\varepsilon \log \varepsilon ).
\label{w3.23}
\end{equation}
For $t\in [\tau +\eta _{\min ,}\eta _{\min }^3],$ we similarly have
\[
|y^\phi -y^{\bar{\phi}}|\leq \int_{\tau +\eta _{\min }}^t\left( (\alpha
+L_5)|y^\phi -y^{\bar{\phi}}|+kL_3C_3||\phi -\bar{\phi}||\right)
ds+C_3||\phi -\bar{\phi}||,
\]
and
\begin{equation}
|y^\phi -y^{\bar{\phi}}|\leq C_4||\phi -\bar{\phi}||,  \label{w3.21}
\end{equation}
where
\begin{equation}
C_4=C_3e^{(\alpha +L_5)(\eta _{\min }^3-\tau -\eta _{\min })}+\dfrac{ e^{(\alpha +L_5)(\eta _{\min }^3-\tau
-\eta _{\min })}-1}{\alpha -L_5} kL_3C_3>C_3.  \label{w3.22}
\end{equation}

Step 5. For $t\in [\eta _{\min }^3,\eta _{\max }^4],$
\[
|y^\phi -y^{\bar{\phi}}|\leq \int_{\eta _{\min }^3}^t\left( ( \delta +
L_3)|y^\phi -y^{\bar{\phi}}|+ kL_4C_4||\phi -\bar{\phi}||\right)
ds+C_4||\phi -\bar{\phi}||.
\]
Thus,
\begin{equation}
|y^\phi -y^{\bar{\phi}}|\leq C_5||\phi -\bar{\phi}||  \label{w3.26}
\end{equation}
where
\begin{equation}
C_5=C_4\left( e^{\delta + L_3(\eta _{\max }^4-\eta _{\min }^3)}+\dfrac{ e^{\delta + L_3(\eta _{\max }^4-\eta
_{\min }^3)}-1}{\delta +L_3}kL_4\right) ..  \label{w3.26a}
\end{equation}

Step 6. For $t\in [\eta _{\max }^4,\tau +\eta _{\max }^4],$ we have
\[
|y^\phi -y^{\bar{\phi}}|\leq \int_{\eta _{\max }}^t\left( (\delta
+L_2)|y^\phi -y^{\bar{\phi}}|+kL_3C_5||\phi -\bar{\phi}||\right)
ds+C_5||\phi -\bar{\phi}||
\]
and
\begin{equation}
|y^\phi -y^{\bar{\phi}}|\leq C_6||\phi -\bar{\phi}||,  \label{w3.27}
\end{equation}
where
\begin{equation}
C_6=C_5(e^{(\delta +L_2)\tau }+\dfrac{e^{(\delta +L_2)\tau }-1}{\delta -L_2} kL_3).  \label{w3.28}
\end{equation}

Step 7. When $t\geq \tau +\eta _{\max }^4,$ both $y$ and $\bar{y}$ are
decreasing and will take the value $1+\eta $ after finite time. Suppose that
$s$ and $\bar{s}$ satisfy
\[
y^\phi (s)=1+\eta ,\;y^{\bar{\phi}}(\bar{s})=1+\eta .
\]
For the later proof, we only consider the case $s<\bar{s},$ since the case
when $s\geq \bar{s}$ can be similarly dealt with and the proof will be
omitted. By Lemma \ref{lemma2} we know
\[
y^\phi (t)=x(t)+O(\varepsilon \log \varepsilon ),\;y^{\bar{\phi} }(t)=x(t)+O(\varepsilon \log \varepsilon ).
\]
By (\ref{x3}), (\ref{x3a}) and (\ref{t5}) we can also obtain
\[
s-(\tau +\eta _{\max }^4)=T_x-(\tau +t_2)+O(\varepsilon \log \varepsilon )
\]
and
\[
\bar{s}-(\tau +\eta _{\max }^4)=T_x-(\tau +t_2)+O(\varepsilon \log
\varepsilon )
\]
where $T_x$ is the period of function $x(t).$ Because the distant between $ \tau +\eta _{\max }^4$ and $s$ may
be greater than $\tau .$ Thus we need split the interval $[\tau +\eta _{\max }^4,s]$ as $[\tau +\eta _{\max
}^4,2\tau +\eta _{\max }^4],$ $[2\tau +\eta _{\max }^4,3\tau +\eta _{\max }^4],\cdots ,[m\tau +\eta _{\max
}^4,s]$ where the length of each interval is exactly $\tau $ except the last one. Here $m$ is the largest
integer less than or equal to $(s-(\tau +\eta _{\max }^4))/\tau .$ We can estimate $ |y^\phi -y^{\bar{\phi}}|$
interval by interval and finally to obtain
\begin{equation}
|y^\phi -y^{\bar{\phi}}|\leq C_7||\phi -\bar{\phi}||,  \label{w3.29}
\end{equation}
with
\begin{equation}
C_7=C_6\left( e^{(\delta +L_2)\tau }+\dfrac{e^{(\delta +L_2)\tau }-1}{\delta
-L_2}kL_2\right) ^{T_x}.  \label{w3.30}
\end{equation}

For $t\in [s,\bar{s}],$ the function $y^{\bar{\phi}}(t)$ satisfies
\[
y^{\bar{\phi}}(t)=1+\eta +O(\varepsilon \log \varepsilon ),\;t\in [s,\bar{s}] \text{ and
}y^{\bar{\phi}}(\bar{s})=1+\eta ,
\]
from which and the equation (\ref{w3.1}) we know that $y^{\bar{\phi}}(t)$ is
decreasing and
\begin{eqnarray*}
|\dfrac{dy^{\bar{\phi}}(t)}{dt}| &=&\left| -(\delta +\beta (y^{\bar{\phi} }))y^{\bar{\phi}}+k\beta (y_\tau
^{\bar{\phi}})y_\tau ^{\bar{\phi}}\right| ,
\\
&=&\left| -\delta (1+\eta )+O(\varepsilon \log \varepsilon )\right| , \\
&\geq &\dfrac{\delta (1+\eta )}2,
\end{eqnarray*}
when $\varepsilon $ is small. Applying the Mean-Value theorem to the
function $y^{\bar{\phi}}(t)$ implies the existence of $\rho \in [s,\bar{s}]$
such that
\[
|y^{\bar{\phi}}(\bar{s})-y^{\bar{\phi}}(s)|=|y^{\bar{\phi}}(\rho )^{\prime }( \bar{s}-s)|\geq \dfrac{\delta
(1+\eta )}2\left| \bar{s}-s\right|
\]
or by (\ref{w3.29}),
\begin{eqnarray}
\left| \bar{s}-s\right| &\leq &\dfrac 2{\delta (1+\eta )}|y^{\bar{\phi}}( \bar{s})-y^{\bar{\phi}}(s)|=\dfrac
2{\delta (1+\eta )}|y^\phi (s)-y^{\bar{
\phi}}(s)|,  \label{idea} \\
&\leq &\dfrac{2C_7}{\delta (1+\eta )}||\phi -\bar{\phi}||.  \nonumber
\end{eqnarray}

Our ultimate goal is give the estimate of $|y_{\bar{s}}^{\bar{\phi}}(\theta
)-y_s^\phi (\theta )|$ where $\theta \in [-\tau ,0].$ Indeed
\begin{equation}
|y_{\bar{s}}^{\bar{\phi}}(\theta )-y_s^\phi (\theta )|\leq |y_{\bar{s}}^{ \bar{\phi}}(\theta
)-y_s^{\bar{\phi}}(\theta )|+|y_s^{\bar{\phi}}(\theta )-y_s^\phi (\theta )|.  \label{wl}
\end{equation}
The first term of the right hand side is bounded by
\[
\int_{s+\theta }^{\bar{s}+\theta }\dfrac{dy^{\bar{\phi}}(t)}{dt}dt\leq
M_2\left| \bar{s}-s\right| .
\]
where $M_2$ is the maximum value of derivative of the function $y^{\bar{\phi} }(t),$ see remark 4.3. The second
term of (\ref{wl}) is bounded by $ C_7||\phi -\bar{\phi}||.$ Thus from (\ref{wl}), we have
\begin{equation}
|y_{\bar{s}}^{\bar{\phi}}(\theta )-y_s^\phi (\theta )|\leq C_7(1+\dfrac{2M_2 }{\delta (1+\eta )})||\phi
-\bar{\phi}||.  \label{w3.31}
\end{equation}
When $\varepsilon \rightarrow 0,$ we have
\[
L_1=O(\dfrac 1{\varepsilon (1+\eta )^{1/\varepsilon
}}),\;\;L_2=O(\varepsilon ),\;\;L_3=O(1/\varepsilon
),\;\;L_4=O(1),\;\;L_5=O(\varepsilon ).
\]
Since $L_1$ is exponentially small as $\varepsilon \rightarrow 0,$ we conclude from (\ref{w3.16}),
(\ref{w3.18}), (\ref{w3.20}), (\ref{w3.22}), ( \ref{w3.26a}), (\ref{w3.28}) and (\ref{w3.30}) that $L_R$ is
exponentially small and satisfies
\[
\lim_{\varepsilon \rightarrow 0}L_R=\lim_{\varepsilon \rightarrow 0}C_7(1+ \dfrac{2M_2}{\delta (1+\eta )})=0.
\]
This completes our proof.\endproof

Since $L_R<1,$ it means that the returning map $R$ is contractive and there
exists a unique fixed-point $\phi $ in $A(\eta )$ so that $R(\phi )=\phi .$
Thus we find a slowly oscillating periodic solution $y(t,\phi )$ for the
equation (\ref{w3.1}). This periodic solution is also globally attractive
for any initial function $\varphi $ in $A(\eta ).$

\section{Full Asymptotic Expansion for the Periodic Solution}

\label{sec:full asymptotic expansion}

In the previous section we use fixed-point theory to prove that there exists
a unique periodic solution in $A(\eta )$ for the equation (\ref{w3.1}) . Now
we like to give a quantitative analysis to this solution. Since the map $R$
is contractive with the Lipschitz constant $L_{R\text{ }}$being
exponentially small, it is possible to give a full asymptotic expansion for
this particular solution.

If we take the initial function $\phi =1+\eta ,$ then we get a solution $ y(t,1+\eta )$ which may not be
periodic. But by Lemma 4.2, $y(t,1+\eta )=x(t)+O(\varepsilon \log \varepsilon )$ when $t$ is finite and there
exists a $T_{1+\eta }>0$ such that
\[
y(T_{1+\eta },1+\eta )=1+\eta ,\;y(T_{1+\eta }+\theta ,1+\eta )\geq 1+\eta
,\theta \in [-\tau ,0).
\]
Then $y(\theta +T_{1+\eta },1+\eta )\in A(\eta ).$

Assume that $y(t)$ is the periodic solution to (\ref{w3.1}) which can be
extended to $(-\infty ,+\infty )$ and satisfies $y(t)\in A(\eta )$ for $t\in
[-\tau ,0].$ Suppose also that for $t\in [-\tau ,0],$ $y(t)$ has the
following asymptotic formula
\begin{equation}
y(t)=\sum_{i=0}^\infty \phi _i(t)  \label{w4.1}
\end{equation}
where $\phi _0(t)=$ $y(t+T_{1+\eta },1+\eta )$ and $\phi _i(t,\varepsilon
),i\geq 1$ will be determined later. Let $y_0(T_0+\theta )$ denote the image
of the returning map of $R$ on $\phi _0,$ i.e.,
\[
y_0(T_0+\theta )=F_\beta (T_0,\phi _0),\text{\ }\theta \in [-\tau ,0]
\]
where $T_0$ $>0$ satisfies
\[
y_0(T_0)=1+\eta ,\;y_0(T_0+\theta )\geq 1+\eta ,\;\theta \in [-\tau ,0).
\]
similarly by induction set
\[
\phi _1(\theta )=y_0(T_0+\theta )-\phi _{0,}\;
\]
\[
y_{i-1}(T_{i-1}+\theta )=F_\beta (T_{i-1},\sum_{j=0}^{i-1}\phi _j)=F_\beta
(T_{i-1},y_{i-2}),\;i\geq 2,
\]
and
\[
\phi _i(\theta )=y_{i-1}(\theta +T_{i-1})-y_{i-2}(\theta
+T_{i-2})_{,}\;i\geq 2,
\]
where for $i\geq 1,$ $T_i>0$ and also satisfies
\[
y_i(T_i)=1+\eta ,\;y_i(T_i+\theta )\geq 1+\eta ,\;\theta \in [-\tau ,0).
\]
Using the result in (\ref{idea}) we can also have
\[
|T_i-T_{i-1}|\leq (L_R)^{i-1}|T_1-T_0|,
\]
which means that the series
\[
T_0+\sum_{j=1}^\infty \left( T_j-T_{j-1}\right)
\]
is absolutely convergent to some constant, say $T_\varepsilon .$ Since by Theorem \ref{theorem2} we have for
$\theta \in [-\tau ,0],$
\begin{eqnarray}
\left| y_i(\theta +T_i)-y_{i-1}(\theta +T_{i-1})\right| &=&\left|
F(T_i,y_{i-1})-F(T_{i-1},y_{i-2})\right|  \label{w4.333} \\
&\leq &L_R\left| y_{i-1}(\theta +T_{i-1})-y_{i-2}(\theta +T_{i-2})\right|
\nonumber \\
&\leq &(L_R)^{i-1}\left| y_1(\theta +T_1)-y_0(\theta +T_0)\right|  \nonumber
\\
&\leq &(L_R)^i|y_0(t,\phi _0)-\phi _0|.  \nonumber
\end{eqnarray}
Then for $\;s\in [-\tau ,0]$, we conclude that
\begin{equation}
y_0(s+T_0)+\sum_{j=1}^\infty \left( y_j(s+T_j)-y_{j-1}(s+T_{j-1})\right)
\label{w4.334}
\end{equation}
is absolutely convergent and have the same value as the initial function $ \sum_{i=0}^\infty \phi _i(s).$ Since
$L_R$ is exponentially small, It is easy to prove that the value of $T_\varepsilon $ is dominated by $T_0$ and
likewise the value of (\ref{w4.334}) is dominated by $y_0(s+T_0),$ each of which has the exponential error
bound. For the leading term $y_0(t),$ obviously we have from Lemma \ref{lemma2} the rough estimates
\[
y_0(t)=x(t)+O(\varepsilon \log \varepsilon ),\;T_0=T_x+O(\varepsilon \log
\varepsilon ).
\]
Next we would like to give the refined estimate for $y_0(t)$ and $T_0$ by
using this information and developing the idea of Lemma \ref{lemma2}.

Like in Lemma \ref{lemma2}, we shall split the interval $[0,T_0]$ and estimate $y_0(t)$ interval by interval.
For illustration, we only need to show the first interval's estimate to the readers. Keep in mind that the
initial data is taken as $y(t+T_{1+\eta },1+\eta )$ which is greater than $ 1+\eta $ when $t$ lies in the
interval $[-\tau ,0).$ Let $\eta _1$ and $\eta _2$ be still the value in Lemma 4.2 and $t_1^{y_0}$ satisfy $
y_0(t_1^{y_0})=1.$ We denote $y_0(t)=y_0(t).$ Integrating the equation (\ref {w3.1}) from $0$ to $t,\;t\in
[0,t_1^{y_0}],$ gives
\begin{eqnarray}
y_0(t)-y_0(0) &=&-\delta \int_0^ty_0(t)dt-\int_0^t\beta (y_0)y_0(t)dt
\label{w5.1} \\
&&+k\int_0^t\beta (y_0(t-\tau ))y_0(t-\tau )dt.  \nonumber
\end{eqnarray}
It is obvious that the last term of the right hand side of (\ref{w5.1}) is
exponentially small and can be viewed as $O(\varepsilon ).$ Next we claim
that
\begin{equation}
\int_0^t\beta (y_0)y_0(t)dt=O(\varepsilon ).  \label{claim}
\end{equation}
To see this, we need to note that $y_0(t)$ is decreasing in $t\in
[0,t_1^{y_0}]$ and $dy_0/dt$ is the order of $O(1)$ or more precisely
\begin{equation}
-\alpha (1+\eta )\leq \dfrac{dy_0}{dt}=-[\beta (y_0)+\delta
]y_0+O(\varepsilon )\leq -\delta +O(\varepsilon ).  \label{w5.2}
\end{equation}
Thus from (\ref{w5.2}) and the fact
\begin{eqnarray*}
\int_0^t\beta (y_0)y_0(t)\dfrac{dy_0}{dt}dt &\leq &\int_0^{t_1^{y_0}}\beta
(y_0)y_0(t)\dfrac{dy_0}{dt}dt, \\
&=&\int_{1+\eta }^1\dfrac{\beta _0y}{1+y^{1/\varepsilon }}dy, \\
&=&O(\varepsilon ),
\end{eqnarray*}
we know that $\int_0^t\beta (y_0)y_0(t)dt$ is also the order of $ O(\varepsilon )$ and the claim of
(\ref{claim}) is true$.$ It follows then from (\ref{w5.1})
\[
y_0(t)\ =-\delta \int_0^ty_0(t)dt+1+\eta +O(\varepsilon ).
\]
Solving this integral equation, we get
\[
y_0(t)=(1+\eta +O(\varepsilon ))e^{-\delta t},
\]
which implies
\begin{equation}
y_0(t)=y_0(t,\varepsilon )=x(t)+O(\varepsilon ),\,t\in [0,t_1^y].
\label{w5.5}
\end{equation}
Using the same approaches above we can prove that in the whole interval $ [0,T_0],$ (\ref{w5.5}) is still true.
Furthermore, we can prove that
\[
T_0=T_x+O\left( \varepsilon \right) ,
\]
which completes our refined estimates.

%-------------------------------------------------------------------------
%%% ---------------------------------------------------------------
\bibliographystyle{siam}
\bibliography{cmlbib}
%%% ----------------------------------------------------------------------

\end{document}